# NUMERICAL MODEL FOR CALCULATING THE FIELD DEPENDENCE OF THE IRREVERSIBLE MAGNETISATION OF HARD SUPERCONDUCTORS IN HIGH PULSED MAGNETIC FIELDS.

J. VANACKEN, L. TRAPPENIERS, K. ROSSEEL, W. BOON

*Laboratorium voor Vaste-Stoffysica en Magnetisme, Katholieke Universiteit Leuven,
Celestijnenlaan 200 D, B-3001 Leuven
E-mail: johan.vanacken@fys.kuleuven.ac.be*

In type II hard superconductors the irreversible magnetization shows an impressive variety of different magnetic field dependencies. In this paper we will try to describe the *M(H)* relation at fixed temperature $T_0$ by a numerical model in which we incorporate two features: the *magnetic* and *electric* field dependency of the local critical current density $j(T_0,H,E)$. The electric field is determined by the magnetic field sweep rate; in effect $E \sim dB/dt$. We confront the model with experimental pulsed field magnetization measurements (PFMM) on a fast melt processed $YBa_2Cu_3O_7$ sample.

## 1 Introduction

The critical parameters of the high temperature superconductors (HTSC) are very often only known in the low field regime ($\mu_0 H_{DC} < 10T$) and therefore the high field values ($\mu_0 H > 30T$) of these parameters must be found from extrapolations. These extrapolations induce some uncertainty and therefore, a direct measurement of the superconducting critical parameters in high fields is still of great importance. To generate these high fields, however, pulsed field techniques are needed. In this way, fields up to $\mu_0 H = 60$ T or more can be obtained, with a typical duration of ~ 20 ms. However, there is one particular feature in measuring superconductors under these conditions: not only the field magnitude, but also the *applied magnetic field sweep rate* ($^{dH}/_{dt}$) is a function of time. The latter is very important, since shielding currents in these materials are sensitive to the $^{dH}/_{dt}$, a consequence of the variation of the voltage V in the voltage-current (I-V) curves [1]. Taking this into account, we modify the Bean model [2] to describe pulsed field magnetisation measurements (PFMM).

## 2 Principle of the calculation

To initiate, we have to define a field versus time pattern; in this pattern, the external applied field values are entered, at equidistant times. In this way, when $dH_{ext}/dt$ is high, the values of successive external fields $H_{ext}(t)$, $H_{ext}(t+\Delta t)$ grow strongly. The sample slab is divided into *2N* different layers parallel to the applied field, and for each layer *i*, the internal local field (in fact induction) is defined as $h(x_i)$. The position parameter $x_i$ is defined by $x_i = i\, d / 2N$ where *d* is the thickness of the sample. At the edge of the sample $i = 0$, whereas at the center of the sample $i = N$. Using the definitions above, the boundary condition (in the thin slab geometry) is simply given by $h(x_0) = H_{ext}$. From this value for *i* (*i=0*) on, we start to check if the difference between the field $h(x_i)$ and $h(x_{i+1})$ is larger than $j(x;T_0,h,E) \cdot \Delta x$, where $\Delta x$ is defined by $\Delta x = x_{i+1}-x_i$. If this would be the case, then at $x_i$ a current larger than $j(x;T_0,h,E)$ would flow, which is not possible. Therefore any rearrangement of the local field gradient has to obey $j|_x < j_c(T_0, H=h(x), E)|_x$. In the calculation, this is incorporated by calculating the local field value at $i+1$ as follows:

$$h(x_{i+1}) = h(x_i) - \text{sgn}[h(x_i) - h(x_{i+1})]\ j(x_i, T_0, h(x_i), E)\ \Delta x$$

After each complete cycle ($i=0$ to $N$), the flux profile at that time $t$ and external field $H_{ext}(t)$ can be visualized. The results of the classical Bean model are reproduced if one uses $j|_x = j_c(T_0)$, $\forall\ x$.

There is no explicit time dependency present in our model, since in pulsed fields, the fast way in which the magnetic field is driven inside the superconductor doesn't allows any substantial influence of relaxation effects.

In this work a separation of variables approximation is made:

$$j(x,t;T_0,h,E) = j_0\ a(T_0)\ b(h)\ g(E)$$

Respectively, the temperature dependence, the magnetic and electric field dependencies are given by $a(T)$, $b(h)$ and $g(E)$. Normally, in an hysteresis measurement, the temperature is fixed at $T=T_0$, which leads to $a(T_0)$ = constant. Moreover, the numerical model can just as easy go beyond the separation of variables approximation, but for demonstrating the numerical model we will use it.

The complete $E$-$J$ characteristic can not be described by one simple function, but in a given region (TAFF, FC, FF, NS), the following relations approximate the $E(j)$ curve in the different $\Delta j$ areas :

$$E = r_i\ (j - j_i)^{m_i} \Rightarrow g(E) = \left[1 + \left(\frac{E}{j_i^m\ r_i}\right)^{1/m_i}\right] \text{ with } E = -\frac{d}{2}\mu_0 \frac{\partial(H+M)}{\partial t}$$

where the indices i represents respectively the TAFF, FC, FF and normal state regime. The constants $j_i$ are the intersecting points at $E = 0$, and describe a kind of an offset current density, since for TAFF, FC and FF the linear (or powerlaw with power $m_i$) does not go through $(E,j) = (0,0)$; in contrast to the normal state resistance. The electric field $E$ was calculated using the Maxwell equations. At this point, we have developed the framework of the numerical model, and we are now able to confront the model with reality.

## 3  Relation between experiments and model.

Since the magnetic field sweep rate in PFMM can be as high as $\mu_0\ dH/dt = 4 \cdot 10^3$ T/s, it is obvious that such an experiment is influenced by dynamic effects. For an oscillating pulse ($\mu_0 H_{max}$ = 7.5 T), the field versus time dependence is below.

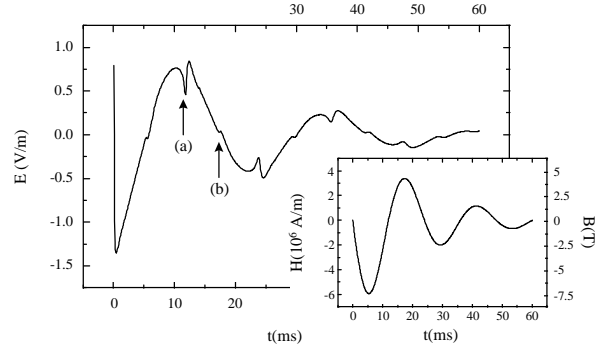

**Figure 1.** Time dependencies of the applied field, sweep rate and induced electric field.

As can be seen from the inset of Fig 1, in an oscillating pulse, $\mu_0 H(t)$ is essentially a decaying sine wave. During the shot, $\mu_0 dH/dt$ changes from $\mu_0 dH/dt \sim 3 \cdot 10^3$ T/s at $B=\mu_0 H \sim 0$ to almost zero at $B \sim B_{max}$. The electric field induced by the changes of the external applied field, and the sample magnetization is shown in figure 1. The overall shape of E is dominated by $dH/dt$ (3$^{th}$ Eq.), and the only effect of $dM/dt$ is observable at (a) and (b), which are respectively the central peak effect and the magnetization polarity switch at maximal fields. In this case, there must be a clear contribution to the magnetic hysteresis loop from the latter. Indeed, as shown in figure 2 the *M(H)* behavior is quite sweep rate dependent and for fixed field, the magnetization values are different. Clearly, the lower $dH/dt$, the lower the apparent magnetization (caused by a lower critical current). Using the equation below, we simulated the experimental curve (inset figure 2). The result of this simulation is shown in figure 2. Several typical features in both graphs are identical, but the fit is not quite perfect.

$$j(x,t;T,h,E) = j_0 \left[ f \exp\left(-h/H_1\right) + (1-f) \exp\left(-h/H_2\right) \right] \left[ 1 + E/j\mathbf{r} \right]$$

The fit parameters are: $j_o = 12.5 \ 10^8$ A/m$^2$, $H_1$=8.5 $10^5$ A/m, $H_2$=2 $10^7$A/m, f=0.75, $1/j\rho$= 0.3 V/m and the size $d$ =5 $10^{-4}$ m. In using the same parameters, the DC hysteresis curve measured by a vibrating sample magnetometer on the same sample can also be fitted. We need to take a combination of *two* exponential decays, to obtain a good fit. Such a combination is well known for granular superconductors, in which the critical current becomes a mixture between the parallel an orthogonal currents due to the large anisotropy of the HTSC materials.

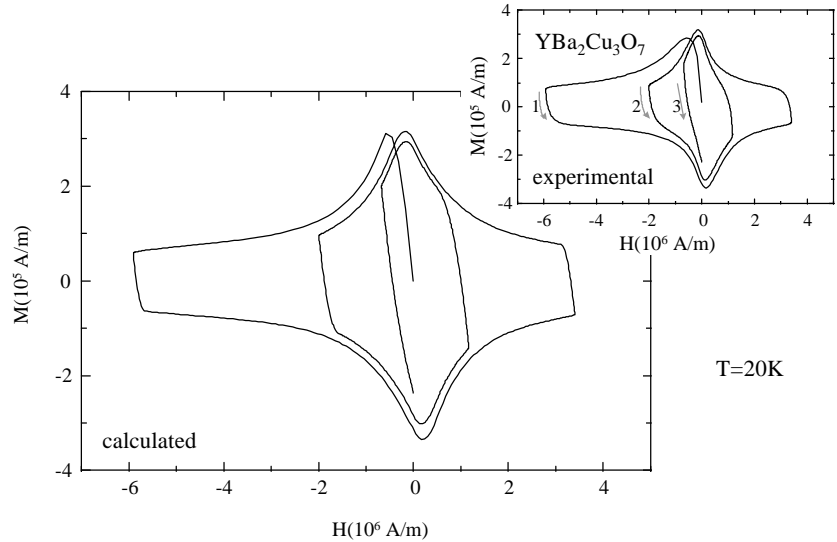

**Figure 2.** Calculated and experimental hysteresis curve for a fast melt processed YBa$_2$Cu$_3$O$_7$ sample

### 4 Conclusions

We have developed a model for calculating the hysteresis loop for hard superconductors in pulsed magnetic fields. Essential is the influence of the electrical field induced by a changing magnetic field and sample magnetization. The presence of an electrical field translates in a higher apparent critical current density where *dH/dt* is large.

### 5 Acknowledgments.

This work is supported by the FWO-Vlaanderen, and the Belgian IUAP programs.

**References.**